\documentclass[11pt, a4paper]{article}

\usepackage{jheppub}

\usepackage{mathrsfs}
\usepackage[T1]{fontenc}
\usepackage{mathpazo}
\usepackage{setspace}
\usepackage{amsfonts}
\usepackage{amssymb}
\usepackage{amsmath}
\usepackage{epsfig}
\usepackage{latexsym}
\usepackage{color}
\usepackage{graphicx}
\usepackage{nicefrac}
\usepackage[latin1]{inputenc}
\usepackage{slashed}
\usepackage{multirow}

\usepackage{hyperref}

\def\be{\begin{equation}}
\def\ee{\end{equation}}
\def\ba{\begin{eqnarray}}
\def\ea{\end{eqnarray}}

\def\del{\partial}

\def\r{\rho}
\def\a{\alpha}

\def\D{\Delta}
\def\e{\epsilon}

\def\C{\Chi}
\def\th{\theta}

\def\s{\sigma}
\def\S{\Sigma}

\def\IR{\relax{\rm I\kern-.18em R}}

\def\inv{^{\raise.0ex\hbox{${\scriptscriptstyle -}$}\kern-.05em 1}}

\def\const{{\rm const.}}

\title{New ${\cal N}=1$ supersymmetric $AdS_5$ backgrounds in Type IIA supergravity }

\author[a]{Konstadinos Sfetsos}
\author[b]{and Daniel C. Thompson}
 \affiliation[a]{Department of Nuclear and Particle Physics\\
Faculty of Physics, University of Athens\\
Athens 15784, Greece}
\affiliation[b]{Theoretische Natuurkunde, Vrije Universiteit Brussel, and
The International Solvay Institutes\\
Pleinlaan 2, B-1050, Brussels, Belgium}

\emailAdd{ksfetsos@phys.uoa.gr}
\emailAdd{Daniel.Thompson@vub.ac.be}

\abstract{We present a family of  ${\cal N}=1$ supersymmetric backgrounds in type-IIA supergravity and their lifts to eleven-dimensional supergravity.
These are of the form  $AdS_5 \times X^5$ and are characterised by an $SU(2)$ structure.
The internal space, $X^5$, is obtained from the known Sasaki-Einstein  manifolds,
$Y^{p,q}$, via an application of non-Abelian T-duality. }


\def\beq{\begin{equation}}
\def\eeq{\end{equation}}
\def\bea{\begin{eqnarray}}
\def\eea{\end{eqnarrat}}
\begin{document}

\def\C{{\cal C}_\zeta}
\def\S{{\cal S}_\zeta}
\def\T{{\cal T}}
\def\R{{\cal R}}

\maketitle
\flushbottom

 \def\baselinestretch{1.2}
 \noindent

\section{Introduction}

According to the AdS/CFT correspondence \cite{Maldacena:1997re},  the ${\cal N}=1$
superconformal field theory arising from a stack of  D3-branes at the tip of a Calabi-Yau cone over an Einstein manifold $X_5$
is conjectured to be dual to the type-IIB supergravity background with metric $AdS_5\times  X_5$  \cite{Kehagias:1998gn,Klebanov:1998hh,Morrison:1998cs,Acharya:1998db}.
The discovery of a countably infinite family of Sasaki--Einstein manifolds known as $Y^{p,q}$ in \cite{Gauntlett:2004yd,Gauntlett:2004hh}
provided a dramatic expansion in concrete examples of the correspondence.
Up to that point, the only explicitly known metrics for $X_5$ were the round metric on $S^5$ and the homogenous space
$T^{1,1}$ -- with corresponding field theories being ${\cal N}=4$ SYM and the ${\cal N}=1$ two-node quiver of Klebanov and Witten \cite{Klebanov:1998hh}.
Following this geometric discovery, and its subsequent interpretation in terms of toric geometry \cite{Martelli:2004wu},
an infinite family of examples of holography could be deduced in a systematic way \cite{Benvenuti:2004dy}.

Non-Abelian T-duality \cite{de la Ossa:1992vc}, the extension of the more familiar T-duality of $U(1)$
isometries to non-Abelian isometry groups,  has been established as a solution generating technique of type-II supergravity backgrounds
supported by Ramond fluxes \cite{Sfetsos:2010uq}.
A natural question, given the importance of Abelian T-duality, to ask is how can this non-Abelian duality be exploited in the context of holography.
Here we will restrict our focus to conformal backgrounds with $AdS_{5}$ factors in the geometries
(a number of related recent works have considered non-Abelian T-duality in various other contexts
\cite{Lozano:2011kb,Itsios:2012dc,Lozano:2012au,Itsios:2012zv,Itsios:2013wd,Jeong:2013jfc,Barranco:2013fza,Gevorgyan:2013xka,Macpherson:2013zba,Lozano:2013oma,Gaillard:2013vsa,Elander:2013jqa,Zacarias:2014wta,Caceres:2014uoa,Pradhan:2014zqa,Lozano:2014ata}).

In  \cite{Sfetsos:2010uq} this application of non-Abelian T-duality was initiated by considering the non-Abelian T-dual
of $AdS_{5}\times S^{5}$ along an $SU(2)$ isometry group acting in the sphere.
The result of this was a solution in type-IIA supergravity that preserved ${\cal N}=2$ supersymmetries that has at least a close relation
to the class of geometries provided by Giaotto and Maldancena \cite{Gaiotto:2009gz} as the gravity duals to Giaotto theories \cite{Gaiotto:2009we}
(in the solution of \cite{Sfetsos:2010uq} there is a singularity making an exact identification difficult at the field theory level).
Following the historical development of type-IIB examples of holography outlined above, a similar non-Abelian T-dualisation was applied in \cite{Itsios:2012zv}
to $AdS_{5}\times T^{1,1}$ again giving rise to geometries in type-IIA but now preserving ${\cal N}=1$ supersymmetry and free from singularities.
Whilst the precise holographic interpretation is not fully understood, the results of \cite{Itsios:2012zv} indicate some connection to the ${\cal N}=1$
counterparts of Giaotto or ``Sicilian'' theories considered in \cite{Bah:2012dg}.

In this paper we report on the result of applying this dualisation to $Y^{p,q}$ spaces.
We will construct an infinite family of smooth solutions to type-IIA supergravity which preserve ${\cal N}=1$
supersymmetry -- these backgrounds can be understood as having an $SU(2)$ structure.  In this short note our aim is to present this geometry which we hope may open the path for a further study of its  holographic  interpretation.

We begin with a telegraphic review of the geometry of $Y^{p,q}$,
then summarise the dualisation procedure and finally give the results of the new geometry.

\section{Some salient features of $Y^{p,q}$}
 We review only the essentials that are needed for our purpose and refer the reader to the original articles
\cite{Gauntlett:2004yd,Gauntlett:2004hh} for further explanations.
The local form of the metric ia
 \be
 \label{Ypqmetric}
 ds^2 = \frac{1-y}{6} \left(\s_1^2 + \s_2^2  \right) + \frac{1}{w(y) v(y)} dy^2  + \frac{v(y)}{9} \s_3^2  + w(y) \left[ d\alpha + f(y) \s_3 \right]^2\ ,
 \ee
 in which the functions are defined by
 \be
    \begin{aligned}
      w(y) = \frac{2 (b - y^2) }{ 1- y} \ , \quad
      v(y) = \frac{b-3y^2 +2 y^3 }{b-y^2} \ ,  \quad
      f(y) = \frac{b-2y + y^2 }{6(b-y^2)} \ ,
    \end{aligned}
 \ee
 where $0<b<1$ is a parameter determined by $p$ and $q$.  The   $\s_{i}$ are a set of $SU(2)$ left invariant one-forms
 \be
\begin{aligned}
 \s_1 = \cos \psi \sin \th d\phi - \sin \psi d\th \ ,   \quad  \s_2 = \sin \psi \sin \th d\phi + \cos \psi d\th \ , \quad    \s_3 = d\psi+ \cos \theta d\phi \ ,
\end{aligned}
\ee
 with the periodicities for the angles
  \be
 0\leqslant \theta \leqslant \pi \ , \quad 0\leqslant \phi \leqslant 2\pi  \ , \quad 0\leqslant \psi \leqslant 2\pi \ .
\ee
The domain of $y$ is
  \be
 y_1 \leqslant y \leqslant y_2\ ,
\ee
  where $y_1$ and $y_2$ are the two smallest roots of
  \be
  b - 3 y^2 + 2 y^3 = 0 \ .
  \ee
  The space $B_{4}$   given by the coordinates $( \theta, \phi, y, \psi )$   is topologically an $S^{2}\times S^{2}$.
The period of the angle $\alpha$ can be fixed, in terms of $p$ and $q$, such that the total space is an $S^{1}$ fibration over $B_{4}$
and is topologically an $S^{2}\times S^{3}$. The isometry group is, up to discrete identifications, $SU(2)\times U(1) \times U(1)$.

 The $Y^{p,q}$ geometries are Sasaki--Einstein meaning that the metric cone
  \be
  ds^2 = dr^2 +  r^2 ds^2(Y^{p,q}) \ ,
  \ee
  is Calabi-Yau and as such is equipped with a K\"ahler form $J$ and a holomorphic three-form $\Omega$.  A natural basis is
   \be
     \begin{aligned}
     \label{Ypqbasis0}
         & \frak{e}^1 = -\frac{1}{\sqrt{6}} H(y)^{-1} dy \quad   \frak{e}^2 = -\frac{1}{\sqrt{6}} H(y) ( d\beta - \cos \theta d\phi )
\quad   \frak{e}^3 =  \frac{\sqrt{1- y}}{\sqrt{6}}  d\theta  \\
         & \frak{e}^4 =  \frac{\sqrt{1- y}}{\sqrt{6}}  \sin \theta d\phi  \quad   \frak{e}^5 =  \frac{1}{3} \left( d\psi + y d\beta
+ (1- y) \cos \theta d\phi \right) \quad \frak{e}^r = d \log r\ ,
    \end{aligned}
 \ee
  where we define $6 H(y)^2 =  w(y) v(y)$ and $\beta= -(6 \alpha +\psi) $.  In this basis,
  \be
    \begin{aligned}
    \label{YpqJOmega}
    &  J  =  r^2 \left( \frak{e}^{r}\wedge \frak{e}^{5} + \frak{e}^{1}\wedge \frak{e}^{2} -\frak{e}^{3}\wedge \frak{e}^{4}  \right) \ ,  \\
    & \Omega_3 = r^3 e^{i \psi}  \left(  \frak{e}^r + i   \frak{e}^5 \right) \wedge  \left( \frak{e}^1
+ i   \frak{e}^2   \right) \wedge   \left( \frak{e}^3 - i  \frak{e}^4   \right)\ ,
    \end{aligned}
  \ee
  where $ \frak{e}^r = d \log r$.  These obey $dJ = 0$ and $d\Omega_3 =0$ and are normalised such that $\Omega_3 \wedge \overline \Omega_3 = \frac{4 i}{3} J\wedge J \wedge J$.

With this space we can construct the following solution of type-IIB supergravity supported by a self-dual RR five-form:
   \be
     \begin{aligned}
        ds^2 &= ds^2[AdS_{5}] + ds^2[Y^{p,q}]  \ , \quad 
          F_5 &= 4 (1 +  \ast)  \frak{e}^1 \wedge \frak{e}^2 \wedge \frak{e}^3 \wedge \frak{e}^4 \wedge \frak{e}^5 \ .
    \end{aligned}
   \ee
   In the basis defined in \eqref{Ypqbasis0} together with the remaining frame-field
on $AdS_5$ specified by $\frak{e}^{x^\mu} = r d x^\mu $ for $\mu = 0 \dots 3$, the  four Killing spinors that do not depend on the $x^\mu$
(and are dual to supersymmetries in the gauge theory) are given by
   \be
\epsilon = e^{- \frac{i}{2} \psi} \sqrt{r} \eta_0\ ,
\ee
where $\eta_0$ is a constant spinor obeying
\be
 i \Gamma_{x^0 x^1 x^2 x^3 }  \eta_0 = - \eta_0 \ , \quad  \Gamma_{12} \eta_0 =  - i \eta_0 \ , \quad \Gamma_{34} \eta_0 = i \eta_0 \  .
\ee
The Killing vectors associated to the $SU(2)$ isometry
\be
\begin{aligned}
\label{su2Killingvec}
    & k_{(1)} = - \cos \phi \partial_{\th} + \cot \th \sin \phi \partial_{\phi}  - \csc \th \sin \phi \partial_\psi  \ , \\
 &  k_{(2)} =- \sin \phi \partial_{\th} - \cot \th \cos \phi \partial_{\phi}  + \csc \th \cos \phi \partial_\psi   \ ,
 \\
    & k_{(3)}=  - \partial_{\phi}  \ ,
\nonumber
\end{aligned}
\ee
may act on the Killing spinor by means of the spinor-Lorentz-Lie derivative
\be
\label{SpinorLorentz}
{\cal L}_k \e = k^\mu ( \partial_\mu +\frac{1}{4} \omega_{\mu AB} \Gamma^{AB}) \e + \frac{1}{4} \nabla_\mu k_\nu \Gamma^{\mu \nu} \e \ . 
\ee
One finds that, by virtue of the projection conditions,
\be
\label{SpinorDerivonYPQ}
{\cal L}_{k_{(i)}} \e =  0 \ , \quad i=1,2,3 \ .
 \ee
This is, of course, no more than the gravity realisation that this $SU(2)$ does not correspond to an R-symmetry in a field theory.
For completeness we note the action of the remaining $U(1)$'s is given by
\be
\begin{aligned}
    &  {\cal L}_{\partial_\psi } \e =  - \frac{i}{2} \epsilon \ ,  \quad  {\cal L}_{\partial_\alpha } \e =  0\ .
\end{aligned}
\ee

\section{Non-Abelian T-duality essentials}
 The most direct way to obtain the relevant transformation rules for the background under non-Abelian T-duality
 is to work with a string $\sigma$-model and follow a Buscher procedure for the NS sector.  We supplement this in the RR sector using the  rules found in \cite{Sfetsos:2010uq}.
 Here we provide a quick summary of this in the absence of ``spectator fields'' ; a comprehensive treatment may be found in \cite{Itsios:2013wd}.

 Consider a  space supporting an isometry group $G$ such that the metric (and the NS two-form if present) can be written  in terms
 of left-invariant Maurer--Cartan forms for this group. Then the $\sigma$-model on this target space is given by
 \be
 S = \int d^2 \sigma E_{ij} L_+^i L_-^j \ ,
 \ee
where  $E_{ij} =  G_{ij}+B_{ij}$ and  $L^i_{\pm}= -i {\rm Tr}(g^{-1} \partial_{\pm} g)$ are the pull-backs of the Maurer--Cartan forms.
We gauge the isometry
\be
\del_\pm  g \to D_\pm g = \del_\pm g - A_\pm g \ ,
\ee
and introduce a  Lagrange multiplier term $-i {\rm Tr}(v F_{+-})$.  Integrating out the Lagrange multipliers enforces a flat connection
and the original $\sigma$-model is recovered upon gauge fixing.  
On the other hand, if we integrate by parts one can solve instead for the non-propagating gauge fields.
If we fix the gauge symmetry (e.g. by setting $g=\mathbb{1}$)   one finds the T-dual sigma model in which the Lagrange
multipliers now play the role of the T-dual coordinates:
\be
\widehat{S}  = \int d^2 \sigma \del_+ v_i  ( M_{ij})^{-1} \del_- v_j  \ , \quad M_{ij}= E_{ij} + f_{ij}{}^k v_k \ .
\label{tdulal}
\ee
Notice that the structure constants and the coordinates themselves explicitly enter into the metric and NS two-form of the T-dual target space
which can be read-off from this sigma model.  The dilaton receives a shift from   integrating out the gauge field and is given by  
  \be\label{TdualDil}
  \widehat{\Phi} = \Phi - \frac{1}{2} \log \det M \ .
  \ee
 As explained in detail in \cite{Sfetsos:2010uq,Itsios:2013wd} this dualisation process acts differently on left and right movers as can be seen in the transformation properties of world-sheet derivatives
 (which actually define a canonical transformation between the two T-dual sigma models \cite{Curtright:1994be,Lozano:1995jx}).
 After T-dualisation, the left and right moving world-sheet bosons each separately define a set of frame-fields for the target space geometry
 which we call $e_{+}^{i}$ and $e_{-}^{i}$.  These are necessarily related by a local frame rotation, $e_{+}^{i} = \Lambda^{i}{}_{j} e_{-}^{j}$,
 with $\Lambda^{T}\Lambda = \mathbb{1}$.   In the example of \eqref{tdulal} this frame rotation has the form $\Lambda = - M^{-T}M$.

 It is this local frame rotation that  determines the action on spinors and the RR sector via the induced spinor representation, $\Omega$,  of $\Lambda$
  \be
  \Omega^{-1} \Gamma^{i} \Omega = \Lambda^{i}{j}    \Gamma^{j}\ .
 \ee
 Consider the RR sector in the democratic formalism that incorporates fluxes and their Hodge duals equally
 (implicitly we are thinking now of a full ten-dimensional type-II context) specified by polyforms
 \be\label{RRpoly}
{\rm IIB}:\   {\mathbb{F}}   =  \sum_{n=0}^4  F_{2n+1}\ ,
\qquad
{\rm IIA}:\   \widehat{\mathbb{F}}  =  \sum_{n=0}^5 F_{2n}\  .
\ee
From these we may construct bi-spinors  $\slashed{\mathbb{F}}$  by contracting the constituent p-forms with p-anti-symmetrised gamma matrices.
The T-daulity rules for the RR sector are then encoded by
 \be
 \label{RRrule}
 e^{\widehat\Phi}   \widehat{\slashed{\mathbb{F}}}   = e^{\Phi} \slashed{\mathbb{F}} \cdot \Omega^{-1}  \ .
 \ee
  One can also view this as a generalisation of a Fourier--Mukai  transformation \cite{Gevorgyan:2013xka}.

Non-Abelian T-duality may or may not preserve supersymmetry if it was there in the original geometry.
A criterion \cite{Sfetsos:2010uq} for the preservation of supersymmetry is that the Killing spinors of the
original geometry  should be invariant under the action given in eq.~\eqref{SpinorLorentz} of the Killing vectors generating the isometry dualised.
In the case at hand, i.e. the dualisation of the $Y^{p,q}$ spaces, we thus anticipate following eq.~\eqref{SpinorDerivonYPQ}
that the supersymmetry will indeed be preserved
in the T-dual.\footnote{Notice that Killing spinor also has a vanishing derivative along  the $U(1)_\alpha$
isometry so a SUSY preserving Abelian T-duality can be performed here, indeed that was part of the duality chain in  \cite{Gauntlett:2004yd,Gauntlett:2004hh}
that led to the discovery of these $Y^{p,q}$ geometries.}
Suppose we start  with ten-dimensional  MW Killing spinors $\epsilon^1$ and $\epsilon^2$, then the Killing spinors in the T-dual will be given by
\be
\label{Tdualspinors}
\hat{\epsilon}^1 = \epsilon^1 \ , \quad \hat{\epsilon}^2  = \Omega\cdot\epsilon^2 \ .
\ee

Instead of working with the explicit Killing spinors, it can be rather convenient in geometries that preserve ${\cal N}=1$ supersymmetry
to work in the language of G-structures,
\cite{Grana:2004bg,Grana:2005sn},  and to harness the power of generalised complex geometry.   The interplay of non-Abelian T-duality
with G-structures was considered in detail in \cite{Barranco:2013fza}.\footnote{Our conventions  essentially follow those of \cite{Martucci:2005ht} as used in \cite{Barranco:2013fza} where further details of the action of non-Abelian T-duality on G-structures can be found.}
In essence one considers a spacetime that is a warped product of four-dimensional Minkowski space and a six-dimensional internal manifold.
Performing an appropriate four-six decomposition of the Killing spinors one is left with internal spinors
$\eta^1_\pm$ and $\eta^2_\pm$ (here the sign denotes six-dimensional chirality). From these one can construct two $Cliff(6,6)$ pure spinors
\be
\Psi_\pm = \eta^1_+ \otimes \eta^2_\pm  \ .
\ee
By means of the Clifford map these can be converted into polyforms. The Gravitino and Dilatino supergravity equations then can be restated as
\be
\label{SUSYeqs}
      e^{-2 A + \Phi} d_H  \left[  e^{2 A - \Phi}  \Psi_1
\right]  =0 \ , \quad
 e^{-2 A + \Phi}  d_H   \left[  e^{2 A - \Phi}  \Psi_2
\right] =
dA \wedge \bar{\Psi}_2
+ \frac{i e^\Phi}{8} \tilde{F}      \ ,
 \ee
where $d_H = d + H\wedge$, $A$ is the warp factor in the metric and $\tilde F$  are the internal components of the RR fluxes in eq.~\eqref{RRpoly}
 i.e. ${\mathbb{F}}  = {\mathbb{G}} + {\rm vol}(R^{1,3}) \wedge    \tilde F$.    For type-IIA one has $\Psi_{1,2} = \Psi_{+,-}$ whereas
 for type-IIB  $\Psi_{1,2} = \Psi_{-,+}$.   An important example occurs when the internal spinors are parallel,
 in which case one finds an $SU(3)$ structure and the pure spinors have the form
\be
\Psi_+ = \frac{i e^A}{8} e^{- i J} \ , \quad \Psi_- = \frac{e^A}{8} \Omega_3\ ,
\ee
where $J$ is a real two-form and $\Omega_3$ a complex three-form.
The $Y^{p,q}$ geometries are of this type with the pure spinors following from \eqref{YpqJOmega}.
When the internal spinors are nowhere parallel, they define an $SU(2)$ structure
\be
\Psi_- = \frac{ e^A}{8} e^{- i j}\wedge z  \ , \quad \Psi_+ = \frac{e^A}{8} e^{\frac{1}{2} z\wedge \bar z} \wedge \omega_2\ ,
\ee
where $j$ is a real two-form, $\omega_2$ a complex two-form and  $z$ a complex one-form.
The action of non-Abelian T-duality on these structures replicates that   of the RR fields  \cite{Barranco:2013fza} namely
\be
\widehat{\slashed{\Psi}}_\pm =  \slashed{\Psi}_\mp \cdot \Omega^{-1} \ .
\ee
 In \cite{Barranco:2013fza} it was shown that under non-Abelian T-duality the $SU(3)$ structure associated with $T^{1,1}$ becomes an $SU(2)$ structure.
 This phenomena is also rather typical of what can happen with Abelian T-duality \cite{Grana:2008yw}
 and we will see that will also be the case for the geometries constructed here.
 (In non-conformal examples of geometries, for instance the Kleabanov-Strassler geometry, it happens that the non-Abelian T-dual
 \cite{Itsios:2013wd,Gaillard:2013vsa}  has what is known as {\em dynamic} $SU(2)$ structure
 where the projections of the pure spinors vary in some directions -- here however the $SU(2)$ structure is of the simpler {\em static} variety).

   \section{The Non-Abelian T-duality Geometry}
   \subsection{Frames and Fields}
   We apply the dualisation proceedure outlined above to $SU(2)$ isometry of the $Y^{p,q}$ solution of type-IIB supergravity.
   To perform the dualisation most efficiently, and to make contact with the general expressions for dualisation
   provided in \cite{Itsios:2013wd}, we choose a set of frame-field for $Y^{p,q}$ given by
    \be
     \begin{aligned}
     \label{Ypqbasis}
        &   \frak{g}^1 = \sqrt{\frac{m(y)}{6}} \s_1   \ , \quad  \frak{g}^2 = \sqrt{\frac{m(y)}{6}} \s_2     \ ,
        \quad   \frak{g}^3 =  \sqrt{g(y)}  \s_3 +  h(y) d\a  \ ,  \\
         &  \frak{g}^\a =  k(y)  d\a     \ , \quad    \frak{g}^y = \frac{1}{\sqrt{v(y) w(y) } } dy     \ ,
    \end{aligned}
 \ee
    in which the supplementary function are defined according to
 \be
 g(y) = \frac{v }{9} + w  f ^2 \ , \qquad h(y) = \frac{w  f  }{\sqrt{g } } \ , \qquad  k(y)^2  =  \frac{ v w  }{9 g} \ , \qquad m(y)= 1-y\ .
 \ee
In this way the T-duality will act only in the $ \frak{g}^{1 \dots 3}$ directions to produce new frame-fields $ \widehat{ \frak{g}}^i_\pm $ in which, as per the discussion in the preceding section,   the $\pm$ denotes the frame-field seen by left and right movers after dualisation.  We fix the  gauge   such that all the Lagrange mulitpliers play the role of T-dual coordinates and parametrize these by 
  \be
   v_1 = \rho \sin(\xi) \ , \quad v_2  = \rho \cos(\xi) \ , \quad  v_3 =  x   \ .
   \ee 
   The target space metric is of the form 
   \be
  \begin{aligned}
   ds^2 &=  ds^2(AdS_5) + ds^2(\widehat{M}_5) \ ,   \\
   ds^2(\widehat{M}_5)  &=    (\frak{g}^\a)^2 +  (  \frak{g}^y)^2 + \sum_{i =1}^3 \widehat{ \frak{g}}^i_\pm \otimes \widehat{\frak{g}}^i_\pm \ ,
     \end{aligned}
   \ee
   with
   \be
  \begin{aligned}\label{eq:frames}
   \widehat{\frak{g}}^1_\pm &=  \frac{2 m^\frac{1}{2} }{\sqrt{3} \Delta} \left( \mp x \rho  \widehat{\sigma}_\pm - \frac{\sqrt{2}}{3}
   \left( 6 \rho^2 + g m \right) d\rho  \right) \ ,\quad 
    \widehat{\frak{g}}^2_\pm  =\frac{  m^\frac{1}{2} }{\sqrt{3} \Delta}  \left( \mp \frac{\sqrt{2}}{3} m \rho \widehat{\sigma}_\pm
    + 4 x g d\rho   \right)   \ ,
    \\
     \widehat{\frak{g}}^3_\pm &= \frac{1}{\Delta} \left( \frac{2m \rho^2}{3\sqrt{g} } \widehat{\sigma}_\pm \mp
     \frac{\Delta}{\sqrt{2g}} dx \mp 4\sqrt{2 g} x \rho d\rho  \right)\ , \quad 
       \widehat{\sigma}_\pm  = 2\sqrt{g} h d\alpha + 2 g d\xi \pm \sqrt{2} d x  \ .
          \end{aligned}
   \ee
   Explicitly one has
   \be
     \begin{aligned}
    \sum_{i =1}^3  \widehat{ \frak{g}}^i_\pm \otimes \widehat{\frak{g}}^i_\pm =& \frac{1}{6 g \Delta}
    \left( (3 \Delta - 4 m \rho^2)dx^2 +  48 g x \rho dx d\rho  + 8 g(m g + 6 \rho^2) d\rho^2 \right. \\
     & \qquad  \left. + 8 g m h^2 \rho^2 d\alpha^2 + 16 m g^\frac{3}{2} h \rho^2 d\alpha d\xi + 8 m g^2 \rho^2 d\xi^2 \right) \ .
               \end{aligned}
    \ee
    Since they define the same metric these frames may be related by a Lorentz transformation $ \widehat{\frak{g}}_+ = \Lambda \widehat{\frak{g}}_-$ whose form may readily be deduced from eq.~\eqref{eq:frames}.    Henceforth we will give all expressions in terms of the {\em plus} frames and no longer write  the corresponding `+' index.
          The dilaton given by eq.~\eqref{TdualDil} is calculated to be
   \be
   \label{eq:dilhat}
   e^{-2\Phi}= \Delta \equiv \frac{2}{9}\left( m(y)^2 g(y) +  6 m(y) \rho^2 +18 g(y) x^2  \right) \ .
   \ee
The NS two-form is given   by
   \be
     \begin{aligned}
   B &= -\frac{\sqrt{6} \rho h  }{ k \sqrt{g} \sqrt{m} }  \frak{g}^\alpha \wedge \widehat{\frak{g}}^2 - \frac{  h  }{ k  }
   \frak{g}^\alpha \wedge \widehat{\frak{g}}^3 - \frac{3\sqrt{2} x }{m}\widehat{\frak{g}}^1  \wedge \widehat{\frak{g}}^2
   - \frac{ \sqrt{6} \rho }{\sqrt{m}\sqrt{g}}\widehat{\frak{g}}^2  \wedge \widehat{\frak{g}}^3  \\
   & =-\frac{ h(3 \Delta - 4 m \rho^2)  }{ 3\sqrt{2 g} \Delta} dx \wedge d\alpha    + \frac{2 \sqrt{2} \r^2 m}{3 \Delta} dx \wedge d\xi\\
   &\qquad +  \frac{4 \sqrt{2} x \r  h g^\frac{1}{2} }{  \Delta}  d\alpha \wedge d\rho +    \frac{4 \sqrt{2} x \r    g   }{  \Delta}  d\xi \wedge d\rho  \ .
     \end{aligned}
     \ee

  The vectorial Lorentz transformation relating the plus and minus frames can be converted to an action on spinors and using  the rule in eq.~\eqref{RRrule} we can  
  deduce the RR fluxes that support this geometry.  This yields
  \be
  \begin{aligned}
  F_2&= -\frac{4}{3}\sqrt{2g} m  \frak{g}^\alpha \wedge  \frak{g}^y =   \frac{4}{9}\sqrt{2} m  dy \wedge d\alpha  \ , \\
  F_4 &= B_2\wedge F_2=  8 x \sqrt{g}    \frak{g}^\alpha \wedge  \frak{g}^y\wedge \widehat{\frak{g}}^1  \wedge \widehat{\frak{g}}^2
  +\frac{8}{\sqrt{3}}\rho \sqrt{m} \frak{g}^\alpha \wedge  \frak{g}^y\wedge \widehat{\frak{g}}^2  \wedge \widehat{\frak{g}}^3 \\
  &=\frac{16}{27\Delta}\rho^{2}m^{2}  dx\wedge dy\wedge d\alpha\wedge d\xi +\frac{32 x\rho g m }{9\Delta} dy\wedge d \alpha\wedge d\xi \wedge d\rho \ .
  \end{aligned}
  \ee
 We have verified using {\tt Mathematica} that this   indeed solves all  the supergravity equations of motion and Bianchi identities.

\subsection{Smoothness}
The expression for the scalar curvature tensor of this background is   unwieldy however there is a
simple test that allows one to see that there should not be any curvature singularities in the T-dual geometry.
In general, if one T-dualises a smooth geometry, then any singularities that occur will do so at a point where the dilaton equally blows up.
This is associated to points in the original geometry where the Killing vectors of the isometry degenerate (for instance at locations where
one dualises a shrinking cycle).   Since the $SU(2)$ Killing vectors have no such points, one anticipates there to be no divergences in the dilaton.
Indeed this is the case;  the dilaton defined in eq.~\eqref{eq:dilhat}  never blows up.
 To see this is the case note that   in terms of $(p,q)$ with $p>q$, the range of $y$ is
   \be\label{yrange}
y_{1}=  \frac{1}{4p} \left(2p - 3q - (4 p^{2} -3 q^{2})^{\frac{1}{2}} \right) < y < y_{2}=  \frac{1}{4p} \left(2p + 3q - (4 p^{2} -3 q^{2})^{\frac{1}{2}} \right)  \ ,
   \ee
   so that the function $m(y)=1-y$ is strictly positive. Then the only possible zeros of $\Delta$ must occur at points where $x=0, \rho=0$ and $g(y)=0$.
   Using the fact that  the parameter $b$ is given by
\be
b= \frac{1}{2} - \frac{1}{4 p^{3}}(p^{2}-3 q^{2}) \sqrt{4p^{2}-3 q^{2}}\ ,
\ee
one can find that the zeros of $g(y)$ occur at
\be
y_{\star} = 1 - \frac{1}{2\sqrt{3} p^{3}} \left( 2p^{6} + p^{5} \sqrt{4p^{2}-3 q^{2}} -3 p^{3} q^{2 } \sqrt{4p^{2}-3 q^{2}}  \right)^{\frac{1}{2}} \ .
\ee
It can be seen that $y_{\star}>y_{2}$ and so any such zeros fall out of the range of the $y$ considered.
Hence the dilaton is nowhere diverging and we thus expect a smooth geometry.  Indeed, one finds explicitly that the scalar curvature is proportional
to (a smooth function times)
\be
 \frac{1}{\Delta^{2} g(y)^{2} m(y)^{2} }
\ee
which remains finite.

  \subsection{$SU(2)$ Structure}
  This geometry preserves the ${\cal N}=1$ supersymmetry of the seed $Y^{p,q}$ background.
  Indeed it can be classified as an (orthogonal, static) $SU(2)$ structure whose pure spinors 
  \be
\Psi_- = \frac{ e^A}{8} e^{- i j}\wedge z  \ , \quad \Psi_+ = \frac{e^A}{8} e^{\frac{1}{2} z\wedge \bar z} \wedge \omega_2 \ ,
\ee
are explicitly given by
\be
\begin{aligned}
z &= \frac{1}{18 \sqrt{\Delta}} \left(  (6 x  - i \sqrt{2} m ) \left[ 2 m r^{-1} dr+3 i\sqrt{2} dx - dy   \right]  + 36 i \sqrt{2} \rho d\rho \right) \\
\omega_{2} &= \frac{\sqrt{2 m} e^{i \xi}}{9 r H \sqrt{\Delta}} \left( 3i \rho dr\wedge dy + r \left[ \rho m dy\wedge d\xi -6y \rho dy\wedge d\alpha -
im dy\wedge d\rho  \right]  \right. \\
&\qquad \qquad \qquad \left. +3 H^{2} \left(6 \rho dr \wedge d\alpha + \rho dr\wedge d\xi - i dr\wedge d\rho - 2 ir \rho d\alpha \wedge d\xi
- 2 r d\alpha \wedge d\rho \right) \right)
\\
j_{2} &= \frac{1}{\Delta} \left( -\frac{1}{18 g r}\left(-v w \Delta + 8 \sqrt{g} h m^{2 } \rho^{2} \right) dr\wedge d\alpha  - \frac{4 m^{2} \rho^{2}}{9 r} dr\wedge d\xi +   \frac{m }{9 g }(\Delta+2 \sqrt{g} h \rho^{2} ) dy\wedge d\alpha \right.\\
&\qquad \left.+\frac{2}{9}m \rho^{2 }  dy\wedge d\xi + \frac{4}{3} \sqrt{g} h m \rho d\alpha \wedge d\rho + \frac{4}{3} g m \rho d\xi \wedge d\rho \right)  \ .
\end{aligned}
\nonumber
\ee
These do indeed solve the susy equations  eq.~\eqref{SUSYeqs}.  We leave further study based on these structures, e.g. finding calibrated cycles and determining scaling dimensions of operators corresponding to wrapped branes for the future.
\subsection{Charges}
We note first that
\be
\begin{aligned}
F_{2}  &= -\frac{4}{9}\sqrt{2} (y-1 ) dy\wedge d\a\ ,
 \\
F_{4}- B_{2} \wedge F_{2}&= 0 \ ,
\\
F_{6}- B_{2} \wedge F_{4} +\frac{1}{2} B_{2}^{2}\wedge F_{2} &= 4 \sqrt{2}  {\rm Vol}(AdS_{5})\wedge (x d x + 2 \rho d\rho) \ ,
 \\
F_{8}- B_{2} \wedge F_{6} +\frac{1}{2} B_{2}^{2}\wedge F_{4}  -\frac{1}{8} B_{2}^{3}\wedge F_{2} &=   8 {\rm Vol}(AdS_{5})\wedge ( \rho d\rho \wedge  d\xi \wedge dx)\ .
\end{aligned}
\ee
 This indicates that the only well-defined and finite Page charge is carried by $D6$ branes and in particular 
  \be
 Q_{Page}^{D6} = \int   -\frac{4}{9}\sqrt{2} (y-1 ) dy \wedge d\a   = \frac{\sqrt{2} \pi q^2  \left( 2p +  (4p^2 - 3 q^2)^\frac{1}{2}  \right)}{3p^2 \left( 3q^2 -2p^2 + p (4p^2 - 3 q^2)^\frac{1}{2}  \right)} = \sqrt{2}\pi {\rm Vol}(Y^{p,q}) , 
 \ee
 in which we used the expressions eq.~\eqref{yrange} and that the period of $\alpha$ is given by
 \be
 l = \frac{q}{3q^2 -2p^2 + p (4p^2 - 3 q^2)^\frac{1}{2} } \ .
 \ee
That the D6 Page charge matches the volume of $Y^{p, q}$ suggests that the D3 charge supporting the $Y^{p,q}$ geometry has been converted entirely to D6 charge.

One potentially useful observation is upon changing to polar coordinates $\rho + i x  = \tilde{r} e^{i \chi}$  then there is a nice cycle on which the B field takes the following form
\be
\{\tilde{r}=\const. \ , y = y_0 \ , \alpha = - \xi  \}\quad \Longrightarrow\quad  B_2 = \frac{\tilde{r}}{\sqrt{2}} \sin\chi d\xi d\chi \ ,
\ee
where $y_0$ is a solution of $h(y) =\sqrt{g(y)}$. One can then readily integrate this quantity over the $S^2$ formed by $(\xi , \chi)$.  Notice that the effect of moving in this $\tilde{r}$ direction can be replicated by large gauge transformations which has similarities to what happens in cascading gauge theories. Similar behaviour was observed in  \cite{Lozano:2013oma} for an $AdS_6$ background  and  in \cite{Lozano:2014ata} in  $AdS_4$ background both obtained via non-Abelian T-duality.  In  \cite{Lozano:2014ata} this was  used to suggest that the apparently non-compact nature of $\tilde{r}$ encodes a  spectral flow in a putative dual $CFT_3$.  One might speculate that a similar interpretation could apply here.

\subsection{M-theory lift}
The lift of the above geometry gives an eleven-dimensional metric
\be
\begin{aligned}
ds^2_{11} &= \Delta^\frac{1}{3}\left(ds^2(AdS_5) + \left[\frac{ dy^2}{ v w } + k^2 d\alpha^2 \right] \right)  \\
 & \qquad + \Delta^{-\frac{2}{3} }\left(\Lambda^2 +  \frac{\rho^2 m}{3 g}\omega^2 +\frac{1}{9}(18 x^2+ m^2) dx^2  + 8 x \rho dx d\rho +  \left( 8\rho^2 +\frac{4}{3} m  \right) d\rho^2  \right)
 \end{aligned}
\ee
where
\be 
\Lambda  = dx^{\#} -  \frac{2 \sqrt{2}  }{9}  (m^2 -1 ) d\alpha   \ , \quad  \omega  = 2  g^\frac{1}{2} h  d\alpha +2 g d\xi \ .
\ee
The four-dimensional space specified by the coordinates $(x,\rho, \zeta,x^{\#})$ is non-trivially fibred over the $(y,\alpha)$ directions due to the functional dependance on the $y$ coordinate in the metric. Because of this, the metric can not be simply cast in the form of  Bah {\em et al} \cite{Bah:2012dg} as was the case for the non-Abelian T-dual of $T^{1,1}$ \cite{Itsios:2012zv,Itsios:2013wd}.  However the solution does preserve ${\cal N}=1$ supersymmetry and so falls under the general classification of  Gauntlett {\em et al} \cite{Gauntlett:2004zh} as we shall now show.

If a solution of eleven-dimensional supergravity of the form
\be
ds^2_{11} = e^\lambda \left( ds^2 (AdS_5) + ds^2 ({\cal M}_6 ) \right)
\ee
preseves ${\cal N}=1$ supersymmetry then   ${\cal M}_6$ admits a local $SU(2)$ structure consisting of the real two-form $J_2$, the complex two-form  $ \Omega_2$ and a pair of real one-forms $K^1, K^2$ which must obey a set of differential equations (eqn's 2.13-2.18 of    \cite{Gauntlett:2004zh}).  These can be obtained from the $SU(2)$ structure of the IIA solution presented above by first lifting to give a seven-dimensional $SU(3)$ structure and then  reducing down to give a six-dimensional $SU(2)$ structure\footnote{Here seven-dimensional refers to the space spanned by the M-theory circle, five internal directions and the AdS radial coordinate and six-dimensional is the same without the radial coordinate.}.  To cast the geometry into the form of  \cite{Gauntlett:2004zh} we  first note that $K^1$ defines a canonical coordinate $Y$ through
\be
K^1= e^{-3\lambda} \sec\zeta d\left[   - \frac{1}{3} m(y) x \right]  = e^{-3\lambda} \sec\zeta dY \ ,
\ee
where
\be
e^\lambda  = \Delta^\frac{1}{6} \ , \qquad \sin\zeta =  \frac{2}{3\Delta^\frac{1}{2} } m(y) x  \ .
\ee
Next we  make a transformation of the angles
\be
\hat{\alpha} = \frac{1}{54}\left( -9 x^{\#}  -2 \sqrt{2} \alpha \right) \, , \quad \hat{\xi} = -6 \alpha - \xi  \, \quad \psi = \xi  \ ,
\ee
such that the coordinate $\psi$ plays a distinguished role as the adapted coordinate for a Killing vector defined by $ \cos\zeta (K^2)^\mu \partial_\mu = -3 \partial_\psi $.
For notation convenience we write $\C= \cos\zeta$ and $\S = \sin \zeta$ and  to simplify the results slightly we perform the coordinate transformation
\be
\hat{\rho} = \rho \sqrt{m(y) g(y)} \ .
\ee
Then we find that
\be
K^2 = - \frac{1}{3} \C d\psi + \frac{1}{9 \C \Delta} \left(6\sqrt{2} m^2 d\hat\alpha + \frac{1}{g} (4\hat  \rho^2 m - 3  g \C^2 \Delta ) d\hat\xi    \right) \ .
\ee
The four-dimensional space
\be
ds^{2}_{4} = ds^2({\cal M}_6 ) - (K^1)^2 - (K^2)^2  = e_{1}^{2}+ e_{1}^{2}+e_{3}^{2} + e_{4}^{2}
\ee
 admits a local $SU(2)$ structure
\be
J_{2} = e_{1}\wedge e_{2}+ e_{3}\wedge e_{4}\ , \quad
\Omega_{2} = e^{i \psi } (e_{1}+ i e_{2})\wedge (e_{3} + i e_{4}) \ ,
\ee
 with frame-field given by
 \be
 \begin{aligned}
 e_{1} &= -\frac{4 \sqrt{2}}{3 \C \Delta\T } \hat\rho m^{3} d\hat\alpha + \frac{2\T \hat\rho}{9\C\Delta g}d\hat\xi \ , \quad
  e_{2}  = \frac{6 \C g}{\T} d\hat\rho + \frac{4 Y \hat\rho m}{\C\D \T } dY \ , \quad
  e_{3}   = -\frac{3\sqrt{2}\R }{\T}d\hat\alpha \ , \\
  e_{4} &= \frac{2  \R \hat \rho}{\T g m }d\hat\rho+ \frac{\T}{2\R g m } dy + \frac{2 Y \R }{\C^{2} T \Delta g m^{4}}\left(6 \hat\rho^{2} m^{2} + 18Y^{2}(9g^{2}- g m^{2}) +g^{2} m^{4} \right) dY
 \end{aligned}
 \ee
in which we define the combinations
\be
\R^{2} = 6 g(9g-m^{2}) \ , \quad \T^{2} = \frac{1}{2}\R^{2}\Delta + \frac{2}{3} g^{2}m^{4}   \ .
\ee
We directly checked that $J,\Omega$ and $K$ defined above satisfy the differential supersymmetry conditions.

This is not quite the end of the story since $dY$ appears explicitly in the above frame-field in addition to the four coordinate differentials $(d\hat\alpha,d\hat\xi,d\hat\rho,dy)$. However $ds_{2}^{4}$ is a metric on a four-dimensional space so such terms need to be removed by a coordinate transformation.   Indeed if we define
 \be\label{eq:FG}
 \hat\rho = F(U,V,Y) \ , \quad y =  G(U,V,Y)
 \ee
 one can see that all the $dY$ terms cancel providing that
\be
\label{eq:FGderivs}
\partial_{Y} F(U, V, Y) =  - \frac{2 Y \hat\rho m}{3 \C^{2 }\Delta g } \ , \quad  \partial_{Y} G(U,V,Y) = -\frac{2Y \R^{2}}{3 \C^{2 }\Delta g m} \ .
\ee
Although we did not find the explicit form of the transformation\footnote{In the case of the non-Abelian T-dual of $T^{1,1}$ the analysis is much easier and this coordinate transformation can be found exactly.  In the notation of \cite{Itsios:2013wd} one finds that the canonical coordinate $Y = \frac{x_{2} }{6}$ and the metric is brought into canonical form with the transformation
\be
54 (x_{1})^{2} = {\rm ProductLog }\left(27 e^{54 (U -36Y^{2})} \right)  \ ,
\ee
where  ${\rm ProductLog }(z)$ is the principal solution of $z=w e^{w}$.  The extra complications in the case at hand are caused by the functional dependance on the $Y^{p,q}$ coordinate $y$  that also gets transformed in eq.~\eqref{eq:FG}.  }, one can readily verify that the metric preserves the complex structure $J_{i}{}^{j}$ and moreover
using eq.~\eqref{eq:FGderivs} that $J_{i}{}^{j}$ is independent of $Y$ as is required.  As is the case with the non-Abelian T-dual of $T^{1,1}$ and the solutions of  \cite{Bah:2012dg}, ${\cal M}_{6}$ is not a complex manifold.

\section{Comments}
Whilst this shows the existence of a wide class of explicit supersymmetric solutions in type-IIA
(and of course their lifts to M-theory) the analysis here has been local in nature.
Establishing the global properties and topology of these solutions will of course be imperative.
Indeed, the understanding of such issues has presented a long standing challenge to non-Abelian duality transformations.
Recent work \cite{Sfetsos:2013wia} (based on works in \cite{Sfetsos:1994vz} and more recently in \cite{Polychronakos:2010hd})
has suggested that, at least in certain circumstances, the non-Abelian T-dual $\sigma$-model can also be
understood as the end point of a flow triggered by a relevant deformation to a certain 2d CFT -- this will shed new light on
the puzzle of the apparent non-compactness of some of the coordinates.

It would, needless to say, be extremely exciting if the geometries presented here can be given a holographic interpretation particularly in the context of ``Sicilian'' gauge theories.
We leave this intriguing question open.

\section*{Acknowledgements}
 We would like to thank G. Itsios and C. Nunez for valuable participation in the early stages of this project and many helpful discussions.  We thank E.O Colgain for correspondence and  Y. Lozano, N. Macpherson, D. Martelli and J. Sparks for comments on an earlier version of this draft.
The research of K.\,Sfetsos is implemented
under the \textsl{ARISTEIA} action (D.654 GGET) of the \textsl{operational
programme education and lifelong learning} and is co-funded by the
European Social Fund (ESF) and National Resources (2007-2013).
  D. Thompson is supported in part by the Belgian Federal Science Policy Office through the Interuniversity
Attraction Pole P7/37, and in part by the ``FWO-Vlaanderen'' through the project G.0114.10N and through
an ``FWO-Vlaanderen'' postdoctoral fellowship project number 1.2.D12.12N.

 \providecommand{\href}[2]{#2}

 \begingroup\raggedright\endgroup


\begin{thebibliography}{99}


\bibitem{Maldacena:1997re}
  J.~M.~Maldacena,
{\em The Large N limit of superconformal field theories and supergravity},
  Adv.\ Theor.\ Math.\ Phys.\  {\bf 2} (1998) 231
     \href{http://arxiv.org/abs/hep-th/9711200}{{\tt hep-th/9711200}}.


\bibitem{Kehagias:1998gn}
  A.~Kehagias,
{\em New type IIB vacua and their F theory interpretation},
  Phys. Lett. {\bf B435} (1998) 337
       \href{http://arxiv.org/abs/hep-th/9805131}{{\tt hep-th/9805131}}.

\bibitem{Klebanov:1998hh}
  I.~R.~Klebanov and E.~Witten,
  {\em Superconformal field theory on three-branes at a Calabi-Yau singularity},
  Nucl. Phys. {\bf B536} (1998) 199
         \href{http://arxiv.org/abs/hep-th/9807080}{{\tt hep-th/9807080}}.

\bibitem{Morrison:1998cs}
  D.~R.~Morrison and M.~R.~Plesser,
{\em Nonspherical horizons. 1.},
  Adv.\ Theor.\ Math.\ Phys.\  {\bf 3} (1999) 1
   \href{http://arxiv.org/abs/hep-th/9810201}{{\tt hep-th/9810201}}.


\bibitem{Acharya:1998db}
  B.~S.~Acharya, J.~M.~Figueroa-O'Farrill, C.~M.~Hull and B.~J.~Spence,
{\em Branes at conical singularities and holography},
  Adv.\ Theor.\ Math.\ Phys.\  {\bf 2} (1999) 1249
     \href{http://arxiv.org/abs/hep-th/9808014}{{\tt hep-th/9808014}}.

\bibitem{Gauntlett:2004yd}
  J.~P.~Gauntlett, D.~Martelli, J.~Sparks and D.~Waldram,
{\em Sasaki-Einstein metrics on $S^2 \times S^3$},
  Adv.\ Theor.\ Math.\ Phys.\  {\bf 8} (2004) 711
     \href{http://arxiv.org/abs/hep-th/0403002}{{\tt hep-th/0403002}}.   

\bibitem{Gauntlett:2004hh}
  J.~P.~Gauntlett, D.~Martelli, J.~F.~Sparks and D.~Waldram,
{\em A New infinite class of Sasaki-Einstein manifolds},
  Adv.\ Theor.\ Math.\ Phys.\  {\bf 8} (2006) 987
       \href{http://arxiv.org/abs/hep-th/0403038}{{\tt hep-th/0403038}}.

\bibitem{Martelli:2004wu}
  D.~Martelli and J.~Sparks,
{\em Toric geometry, Sasaki-Einstein manifolds and a new infinite class of AdS/CFT duals},
  Commun.\ Math.\ Phys.\  {\bf 262} (2006) 51
       \href{http://arxiv.org/abs/hep-th/0411238}{{\tt hep-th/0411238}}.


\bibitem{Benvenuti:2004dy}
  S.~Benvenuti, S.~Franco, A.~Hanany, D.~Martelli and J.~Sparks,
{\em An Infinite family of superconformal quiver gauge theories with Sasaki-Einstein duals},
  JHEP {\bf 0506} (2005) 064
         \href{http://arxiv.org/abs/hep-th/0411264}{{\tt hep-th/0411264}}.



\bibitem{Gauntlett:2004zh}
  J.~P.~Gauntlett, D.~Martelli, J.~Sparks and D.~Waldram,
{\em Supersymmetric AdS(5) solutions of M theory},
  Class.\ Quant.\ Grav.\  {\bf 21} (2004) 4335
           \href{http://arxiv.org/abs/hep-th/0402153}{{\tt hep-th/0402153}}.

\bibitem{de la Ossa:1992vc}
  X.~C.~de la Ossa and F.~Quevedo,
  {\em Duality symmetries from nonAbelian isometries in string theory},
  Nucl. Phys. {\bf B403} (1993) 377
        \href{http://arxiv.org/abs/hep-th/9210021}{{\tt hep-th/9210021}}.

\bibitem{Sfetsos:2010uq}
  K.~Sfetsos and D.~C.~Thompson,
{\em On non-abelian T-dual geometries with Ramond fluxes},
  Nucl.\ Phys. {\bf B846} (2011) 21
    \href{http://arxiv.org/abs/1012.1320}{{\tt arXiv:1012.1320}}.


\bibitem{Lozano:2011kb}
  Y.~Lozano, E.~.O Colgain, K.~Sfetsos and D.~C.~Thompson,
{\em Non-abelian T-duality, Ramond Fields and Coset Geometries},
  JHEP {\bf 1106} (2011) 106
  \href{http://arxiv.org/abs/1104.5196}{{\tt arXiv:1104.5196}}.



\bibitem{Itsios:2012dc}
  G.~Itsios, Y.~Lozano, E.~.O Colgain and K.~Sfetsos,
{\em Non-Abelian T-duality and consistent truncations in type-II supergravity},
  JHEP {\bf 1208} (2012) 132
  \href{http://arxiv.org/abs/1205.2274}{{\tt arXiv:1205.2274}}.



\bibitem{Lozano:2012au}
  Y.~Lozano, E.~O Colgain, D.~Rodriguez-Gomez and K.~Sfetsos,
{\em Supersymmetric $AdS_6$ via T Duality},
  Phys.\ Rev.\ Lett.\  {\bf 110} (2013) 23,  231601
     \href{http://arxiv.org/abs/1212.1043}{{\tt arXiv:1212.1043}}.
\cite{Itsios:2012zv}

\bibitem{Itsios:2012zv}
  G.~Itsios, C.~Nunez, K.~Sfetsos and D.~C.~Thompson,
{\em On Non-Abelian T-Duality and new N=1 backgrounds},
  Phys. Lett. {\bf B721} (2013) 342
          \href{http://arxiv.org/abs/1212.4840}{{\tt arXiv:1212.4840}}.



\bibitem{Itsios:2013wd}
  G.~Itsios, C.~Nunez, K.~Sfetsos and D.~C.~Thompson,
 {\em Non-Abelian T-duality and the AdS/CFT correspondence:new N=1 backgrounds},
   Nucl.\ Phys. {\bf B873} (2013) 1
        \href{http://arxiv.org/abs/1301.6755}{{\tt arXiv:1301.6755}}.



\bibitem{Jeong:2013jfc}
  J.~Jeong, O.~Kelekci and E.~O Colgain,
{\em An alternative IIB embedding of F(4) gauged supergravity},
  JHEP {\bf 1305} (2013) 079
      \href{http://arxiv.org/abs/1302.2105}{{\tt arXiv:1302.2105}}.



\bibitem{Barranco:2013fza}
  A.~Barranco, J.~Gaillard, N.~T.~Macpherson, C.~Nunez and D.~C.~Thompson,
{\em G-structures and Flavouring non-Abelian T-duality},
  JHEP {\bf 1308} (2013) 018
    \href{http://arxiv.org/abs/1305.7229}{{\tt arXiv:1305.7229}}.



\bibitem{Gevorgyan:2013xka}
  E.~Gevorgyan and G.~Sarkissian,
{\em Defects, Non-abelian T-duality, and the Fourier-Mukai transform of the Ramond-Ramond fields},
  JHEP {\bf 1403} (2014) 035
    \href{http://arxiv.org/abs/1310.1264}{{\tt arXiv:1310.1264}}.

\bibitem{Macpherson:2013zba}
  N.~T.~Macpherson,
  {\em Non-Abelian T-duality, $G_2$-structure rotation and holographic duals of $N=1$ Chern-Simons theories},
  JHEP {\bf 1311} (2013) 137
   \href{http://arxiv.org/abs/1310.1609}{{\tt arXiv:1310.1609}}.


\bibitem{Lozano:2013oma}
  Y.~Lozano, E.~O Colgain and D.~Rodriguez-Gomez,
  {\em Hints of 5d Fixed Point Theories from Non-Abelian T-duality},
  JHEP {\bf 1405} (2014) 009
  \href{http://arxiv.org/abs/1311.4842}{{\tt arXiv:1311.4842}}.


\bibitem{Gaillard:2013vsa}
  J.~Gaillard, N.~T.~Macpherson, C.~Nunez and D.~C.~Thompson,
{\em Dualising the Baryonic Branch: Dynamic SU(2) and confining backgrounds in IIA},
  Nucl.\ Phys. {\bf B884} (2014) 696
     \href{http://arxiv.org/abs/1312.4945}{{\tt arXiv:1312.4945}}.


\bibitem{Elander:2013jqa}
  D.~Elander, A.~F.~Faedo, C.~Hoyos, D.~Mateos and M.~Piai,
{\em Multiscale confining dynamics from holographic RG flows},
  JHEP {\bf 1405} (2014) 003
   \href{http://arxiv.org/abs/1312.7160}{{\tt arXiv:1312.7160}}.


\bibitem{Zacarias:2014wta}
  S.~Zacarias,
{\em Semiclassical strings and Non-Abelian T-duality},
 \href{http://arxiv.org/abs/1401.7618}{{\tt arXiv:1401.7618}}.


\bibitem{Caceres:2014uoa}
  E.~Caceres, N.~T.~Macpherson and C.~Nunez,
 {\em New Type IIB Backgrounds and Aspects of Their Field Theory Duals},
     \href{http://arxiv.org/abs/1402.3294}{{\tt arXiv:1402.3294}}.

\bibitem{Pradhan:2014zqa}
  P.~M.~Pradhan,
{\em Oscillating Strings and Non-Abelian T-dual Klebanov-Witten Background},
           \href{http://arxiv.org/abs/1406.2152}{{\tt arXiv:1406.2152}}.

\bibitem{Lozano:2014ata}
  Y.~Lozano and N.~T.~Macpherson,
{\em A new $AdS_4/CFT_3$ Dual with Extended SUSY and a Spectral Flow},
           \href{http://arxiv.org/abs/1408.0912}{{\tt arXiv:1408.0912}}.



\bibitem{Gaiotto:2009we}
  D.~Gaiotto,
{\em N=2 dualities},
  JHEP {\bf 1208} (2012) 034
        \href{http://arxiv.org/abs/0904.2715}{{\tt arXiv:0904.2715}}.

\bibitem{Gaiotto:2009gz}
  D.~Gaiotto and J.~Maldacena,
{\em The Gravity duals of N=2 superconformal field theories},
  JHEP {\bf 1210} (2012) 189
        \href{http://arxiv.org/abs/0904.4466}{{\tt arXiv:0904.4466}}.

\bibitem{Bah:2012dg}
  I.~Bah, C.~Beem, N.~Bobev and B.~Wecht,
 {\em Four-Dimensional SCFTs from M5-Branes},
  JHEP {\bf 1206} (2012) 005
      \href{http://arxiv.org/abs/1203.0303}{{\tt arXiv:1203.0303}}.




\bibitem{Curtright:1994be}
  T.~Curtright and C.~K.~Zachos,
{\em Currents, charges, and canonical structure of pseudodual chiral models},
  Phys.\ Rev. {\bf D49} (1994) 5408
  \href{http://arxiv.org/abs/hep-th/9401006}{{\tt hep-th/9401006}}.


\bibitem{Lozano:1995jx}
  Y.~Lozano,
  {\em NonAbelian duality and canonical transformations},
Phys.\ Lett.\ B {\bf 355} \href{http://arxiv.org/abs/hep-th/9503045}{{\tt hep-th/9503045}}.





\bibitem{Grana:2004bg}
  M.~Grana, R.~Minasian, M.~Petrini and A.~Tomasiello,
  {\em Supersymmetric backgrounds from generalized Calabi-Yau manifolds},
  JHEP {\bf 0408} (2004) 046
Phys. Lett. {\bf B355} \href{http://arxiv.org/abs/hep-th/0406137}{{\tt hep-th/0406137}}.


\bibitem{Grana:2005sn}
  M.~Grana, R.~Minasian, M.~Petrini and A.~Tomasiello,
  {\em Generalized structures of N=1 vacua},
  JHEP {\bf 0511}, 020 (2005)
   \href{http://arxiv.org/abs/hep-th/0505212}{{\tt hep-th/0505212}}.


\bibitem{Grana:2008yw}
  M.~Grana, R.~Minasian, M.~Petrini and D.~Waldram,
  {\em T-duality, Generalized Geometry and Non-Geometric Backgrounds},
  JHEP {\bf 0904} (2009) 075
           \href{http://arxiv.org/abs/0807.4527}{{\tt arXiv:0807.4527}}.

\bibitem{Martucci:2005ht}
  L.~Martucci and P.~Smyth,
{\em Supersymmetric D-branes and calibrations on general N=1 backgrounds},
  JHEP {\bf 0511} (2005) 048
     \href{http://arxiv.org/abs/hep-th/0507099}{{\tt hep-th/0507099}}.

\bibitem{Sfetsos:2013wia}
  K.~Sfetsos,
{\em Integrable interpolations: From exact CFTs to non-Abelian T-duals},
  Nucl. Phys. {\bf B880} (2014) 225
      \href{http://arxiv.org/abs/1312.4560}{{\tt arXiv:1312.4560}}.

\bibitem{Sfetsos:1994vz}
  K.~Sfetsos,
{\em Gauged WZW models and nonAbelian duality},
  Phys. Rev. {\bf D50} (1994) 2784
   \href{http://arxiv.org/abs/hep-th/9402031}{{\tt hep-th/9402031}}.

\bibitem{Polychronakos:2010hd}
  A.~P.~Polychronakos and K.~Sfetsos,
{\em High spin limits and non-abelian T-duality},
  Nucl. Phys. {\bf B843} (2011) 344
     \href{http://arxiv.org/abs/1008.3909}{{\tt arXiv:1008.3909}}.

\end{thebibliography}
\end{document}